# Label-free detection of *Giardia lamblia* cysts using a deep learning-enabled portable imaging flow cytometer

Zoltán Göröcs[1,2,3], David Baum[1], Fang Song[1], Kevin DeHaan[1,2,3], Hatice Ceylan Koydemir[1,2,3], Yunzhe Qiu[1], Zilin Cai[1], Thamira Skandakumar[2], Spencer Peterman[2], Miu Tamamitsu[1], and Aydogan Ozcan[1,2,3,*]

[1]Electrical and Computer Engineering Department, [2]Bioengineering Department, and [3]California NanoSystems Institute (CNSI)
University of California, Los Angeles, CA, 90095, USA

*ozcan@ucla.edu

**Abstract**
We report a field-portable and cost-effective imaging flow cytometer that uses deep learning to accurately detect *Giardia lamblia* cysts in water samples at a volumetric throughput of 100 mL/h. This flow cytometer uses lensfree color holographic imaging to capture and reconstruct phase and intensity images of microscopic objects in a continuously flowing sample, and automatically identifies *Giardia Lamblia* cysts in real-time without the use of any labels or fluorophores. The imaging flow cytometer is housed in an environmentally-sealed enclosure with dimensions of 19 cm × 19 cm × 16 cm and weighs 1.6 kg. We demonstrate that this portable imaging flow cytometer coupled to a laptop computer can detect and quantify, in real-time, low levels of *Giardia* contamination (e.g., <10 cysts per 50 mL) in both freshwater and seawater samples. The field-portable and label-free nature of this method has the potential to allow rapid and automated screening of drinking water supplies in resource limited settings in order to detect waterborne parasites and monitor the integrity of the filters used for water treatment.

**Introduction**

Waterborne diseases are caused by the presence of pathogens in drinking water due to the lack of filtration or improper sanitation. *Giardia lamblia* is a microscopic waterborne parasite that affects 200 million people every year with the diarrheal illness known as giardiasis[1]. This parasitic microorganism is found in the feces of infected animals or humans, and can spread through contaminated soil, water or food. *Giardia* takes two forms during its lifecycle: trophozoite and cyst. The trophozoites of *Giardia lamblia* are motile and multiply in the intestines of the host to produce the oval-shaped cyst form of the parasite. The motile form of the parasite cannot survive in the environment, but the cyst form can survive for >2 months in water temperatures of less than 10 °C[2]. If a water sample contaminated with *Giardia* cysts enters the body of a host, the cysts will grow to continue their life cycle. Within 3 to 25 days of the ingestion of cysts, symptoms such as abdominal cramps and bloating begin to appear.

There have been major efforts by federal agencies and governments to reduce the risk of giardiasis by using systems to filter or disinfect water sources (see e.g., Environmental Protection Agency (EPA)'s Surface Water Treatment Rules (SWTRs))[3]. However, there is still a need to detect and count *Giardia* cysts in water samples to ensure proper operation of such filter systems and adjust the level of disinfecting agents for water treatment. For example, EPA Method 1623.1 is focused on the detection of *Cryptosporidium* and *Giardia* in water samples by using filtration, immunomagnetic separation, and fluorescence microscopic analysis. This EPA method has been widely adopted and enables the detection of *Giardia lamblia* cysts in water samples with high specificity and sensitivity. However, this method requires a laboratory infrastructure to perform the measurement as it requires *Giardia* specific fluorescence labels and a microbiology expert to operate the fluorescence microscope and perform analysis. There have been a variety of *Giardia* detection/identification methods[4] reported in the literature to eliminate these drawbacks such as smartphone-based microscopy[5], spectroscopy[6], molecular methods (e.g. PCR[7] and LAMP[8]), flow cytometry[9–11], and laser scanning cytometry[12]. Despite providing some unique advantages such as high sensitivity and specificity, these approaches still require *Giardia* specific chemical labels and related sample preparation methods that require a lab infrastructure.

Here we report a portable imaging flow cytometer that is capable of detecting and enumerating *Giardia Lamblia* cysts in water samples in real-time and without the use of any labels (see Figure 1). This portable imaging flow cytometer analyzes continuously flowing water samples at a throughput of 100 mL/h and is based on lensfree on-chip holographic microscopy.[13–16] The system uses a small peristaltic pump to continuously flow the water sample through a 5 mm wide and 0.8 mm tall rectangular channel, which is placed in direct contact with the image sensor chip without any optical components in between. Synchronized 120 μs long light pulses from a red, green and blue LED provide partially coherent illumination on the sample volume, and generate the holographic signatures of all the objects in a sample volume of 24 μL (6.5 mm × 4.6 mm × 0.8 mm) per each captured image frame. As the objects are flowing through the channel, these captured holograms are reconstructed in real-time using a laptop computer to obtain their color intensity and phase images. These reconstructed holographic images are then automatically classified by a convolutional neural network, and the images containing *Giardia Lamblia* cysts are digitally sorted and counted in real-time (Figure 2). This system weighs 1.6 kg and is field-portable, spanning 19 cm × 19 cm × 16 cm; it is also cost effective with a total cost of <$2500, excluding the laptop computer used to control the system.

We experimentally demonstrate that our imaging flow cytometer can accurately identify water samples contaminated by *Giardia lamblia* cysts with a detection limit better than 10 cysts/50 mL, with the test/imaging taking ~30 min without the use of any labeling or sample preparation. We also demonstrate that this imaging flow cytometer system is compatible with mechanical filtration based pre-concentration techniques, and can be used to analyze 1 US Gallon water samples containing *Giardia lamblia* cysts within ~1 hour. This field-portable and label-free imaging cytometer can especially be useful to monitor the quality of water purification systems and treatment plants[17,18] by providing automated *Giardia* cyst detection without the use of any chemicals and sample preparation.

**Results & Discussion**

The presented computational imaging flow cytometer achieves label-free *Giardia lamblia* cyst detection and quantification by performing deep neural network-based classification of the acquired phase and intensity images of each microscopic object within the water sample under test. Unlike other established methods, there is no labeling or related sample preparation needed to provide specificity to the detection process. Since this imaging flow cytometer analyzes the water sample of interest nondestructively, without contaminating the sample with chemicals (Fig. 2), it also allows subsequent testing of the measured sample with alternative methods that can be used for comparison and validation. This is especially important as we observed a low and unstable rate of *Giardia* cyst recovery from the manufacturer provided *Giardia* sample vials, and therefore comparison to the manufacturer's reported nominal cyst count in the sample was not practical. Instead, after each experiment we collected the effluent of the imaging flow cytometer on a 0.45 µm pore size filter paper. Fluorescence staining and benchtop microscopy were then used to provide gold standard cyst counts, which were compared with each one of our measurements for validation of our label-free test results (see Methods). Accordingly, we tested our system with 50 mL water samples spiked with *Giardia cysts*: as shown in Figure 3, the cytometer correctly identified all the spiked samples as positive (even the samples spiked with less than 10 cysts per 50 mL), all the control samples as negative, and exhibited a ~83% recovery rate on counting the number of *Giardia* cysts within the sample, which is approximately 3-fold higher than the EPA requirement (>27%).[19] Since *Giardia lamblia* can also be transmitted by recreational swimming in both freshwater and marine waters that are contaminated by raw sewage[20–22], next we explored the applicability of our flow cytometer to detect *Giardia* cysts in water samples with higher salinity. For this purpose, we tested *Giardia* spiked seawater samples and achieved a cyst recovery rate of 74% (see Figure 4).

We also investigated the measurement repeatability of our system (Figure 5). For these repeatability tests, we prepared a 50 mL reagent grade water sample, spiked with several thousand *Giardia* cysts, and repeatedly measured it, i.e., the imaging flow cytometer was used to test the same sample three times by collecting and re-testing the effluent. While the resulting numbers of *Giardia* and non-*Giardia* particles detected by our system were similar in all three measurements, some particle loss in the subsequent measurements was observed, most likely due to particles adhering to the sample container and thus disappearing from the solution. This effect was relatively more pronounced for the *Giardia* cyst counts, which might be due to the known preference *Giardia* cysts to adhere to hydrophobic surfaces like the walls of the disposable tubes that were used to hold the water samples during our experiments[23].

Next we investigated the use of pre-filtration to concentrate the water sample of interest and thus increase the overall throughput of each measurement. For this, we spiked bottles of drinking water with *Giardia* cysts (each 1 US gallon containing ~1,200 cysts), rapidly filtered each gallon though a membrane with a pore size of 0.45 µm, and then re-suspended the filtered particles into 50 mL test samples that were automatically screened by our imaging flow cytometer (see Methods section for details). This pre-filtration step also introduces non-*Giardia* particles from the filter into the concentrated volume. We repeated these experiments with three independent positive samples (each 1 US Gallon) and our system was successful in identifying all three samples as positive, and the control sample as negative (see Table 1). This pre-concentration step is performed in ~30 min, and allows the presented system to screen ~1 US Gallon of water within ~1 hour (30 min for imaging and 30 min for pre-filtration) for the presence of *Giardia* cysts.

Different from existing fluorescence based *Giardia* cyst detection methods, which rely on fluorescence staining for the detection and identification of the cysts, our imaging method is label-free, and solely uses the captured phase and intensity images of the particles to perform the sensing task. An important advantage provided by our technique is its ability to measure the flowing water sample in real-time and undisturbed/unprocessed, as no chemicals are added and no incubation time is needed. This technology allows for continuous, cost-effective, and automated screening of water samples for *Giardia* cyst contamination. In order to perform this detection and classification task automatically, we trained a convolutional neural network (Methods) using labeled image datasets of *Giardia* cysts as well as other non-*Giardia* particles, with each set containing tens of thousands of objects. Imaging of these thousands of objects for training the neural network can be accomplished in a single run of our cytometer due to its high throughput. However, to provide a more diverse training image dataset, accounting for e.g., the salinity of the sample and potential variations in the imaging conditions, measurements taken over several days were used to train the neural network. Creating a labeled dataset containing the phase and intensity images of *Giardia* cysts, however, is relatively challenging since some non-*Giardia* particles can look similar to *Giardia* cysts, thus manual labeling can be noisy. Generating a training water sample that contains only *Giardia* cysts is also difficult, as even filtered water samples will collect some other particles from the air and the sample holders used. Therefore, during the *training phase* we spiked each 50 mL filtered water sample to be measured with a high concentration (>1000 cyst/mL) of *Giardia* cysts in order to reduce the ratio of non-*Giardia* particles within the sample to be <5% of the detected objects. This image dataset was further processed to remove non-*Giardia* particles based on their size and shape, improving the purity of our training labels. To allow real-time processing of the imaged particles using a laptop computer, we selected a fairly compact neural network model (DenseNet-121)[24] to perform rapid detection of target cysts, which may be further improved in terms of its specificity using larger/deeper networks, at the cost of additional computation time.

During the training of the classification neural network, we prioritized minimizing the likelihood of false positives. This was necessary as in a practical scenario, each water sample under test (including bottled drinking water samples) will naturally contain thousands of non-*Giardia* particles and misclassifying even a small portion of them would make the whole sample test positive (false) for *Giardia* contamination. Our trained neural network has an average probability of 0.25% to misclassify a non-*Giardia* particle as a cyst, which we have digitally compensated for by subtracting 0.5% of the number of total detected/counted particles in a given measurement from the final *Giardia* cyst count. This offset subtraction procedure helps us avoid false positive decisions at the sample level and can be fine-tuned based on the type of water under test. We should also note that the presented method can be used along with some of the existing sample concentration methods, such as immunomagnetic separation[25], which might further help with its specificity and detection efficiency.

Using the computational power of the Nvidia GTX 2080 GPU installed on the controlling laptop computer, the presented field-portable imaging flow cytometer, on average, can read in the data to the GPU, perform background subtraction, and automatically segment objects from a single full field of view hologram, all within ~34 ms, which is a one-time effort for the entire image FOV. During the processing of the individual segmented sub-images of a given FOV, our program can autofocus on a single object in ~7 ms, perform high resolution color reconstruction of a detected object in 2.5 ms, determine its size in ~6 ms, and finally achieve *Giardia* detection/classification using the trained convolutional neural network in ~9 ms, making the total time to reconstruct, measure, and classify a single object <30 ms. Our portable imaging cytometer captures holograms at a rate of three frames per second, which allows ~10 objects per field of view to be processed in real-time, which corresponds to a maximum of ~1000 objects/1mL for real-time processing at a flowrate of 100 mL/h. For larger sample concentrations (>1000 particles/mL), the device can either save the raw sensor data during the experiments to process it later or real-time operation is also possible at a reduced flowrate and sample throughput. However, this does not set a practical limitation since the statistically significant number of detection events does not change between dense and sparse samples; for denser samples with >1000 particles/mL, by reducing the flowrate one can still accumulate enough particle statistics within the same total test time that is used under lower density (<1000 particles/mL) samples since the total processing time per particle is the same in both cases (<30 ms). In fact, our system permits dynamic control and adjustment of the flowrate on demand, which can be automatically adjusted based on the density of the particles for a given water sample under test. The maximum sample concentration that can be processed by this system is by and large determined by the speed of the used GPU, and therefore the emerging generation of faster GPUs will automatically improve the performance of this imaging flow cytometer, also allowing larger, and potentially more accurate networks to perform cyst classification in real-time.

**Materials and Methods**

*Imaging flow cytometer hardware*

Our imaging flow cytometer uses a 14 Megapixel color CMOS camera with a pixel size of 1.4 µm (Basler aca4600-10uc) to capture lensfree holograms of flowing objects. To gain access to the image sensor, we removed the camera's original housing and replaced it with a fan cooled copper housing to reduce the heating of the camera board. The new housing also serves as a secure holder for the flow channel placed in direct contact with the image sensor's cover glass. The illumination is provided by an RGB LED (Ledengin LZ4-04MDPB), which is filtered by a pair of triple bandpass filters (Edmund Optics #87-246, Chroma Inc. 69015m), and then reflected from a convex mirror (Edmund Optics #64-061) to adjust its temporal and spatial coherence, respectively. The resulting illumination has its center wavelengths at 450nm, 530 nm and 630 nm, and a bandwidth of <10 nm per each color channel, respectively. All the illumination wavelengths are emitted *simultaneously* in 120 µs pulses to allow the capture of sharp holograms from continuously moving objects. The device is powered with 5V, which can be provided by a wall mount power supply or external (e.g., smartphone) batteries. A capacitor charge controller (LT3750, Linear Technologies) is charging three 0.1F capacitors to ~12V, which are used to supply the charge needed for the high current LED pulses. The LEDs are driven by a triple output LED controller (LT3797, Linear Technologies), which is triggered by the image sensor's flash output signal to provide constant current pulses to the LEDs. The drive current of each LED is set independently by the controller software using i2c controlled potentiometers (TPL0401A, Texas Instruments) to achieve adequate white balance. The PCB also integrates a control circuit to power and control a small peristaltic pump (Instech p625) facilitating the sample flow. The speed of the pump, and thus the sample flowrate can also be controlled by the software. An Arduino microcontroller (TinyDuino, Tinycircuits) provides an interface for i2c communications between the device and the controlling laptop. The power consumption of the device during a measurement is under 5 W, which includes the image sensor's typical power consumption of ~2.8 W.

*Hologram reconstruction*

During the measurement of a sample at a flowrate of 100 mL/h the image sensor inside the flow cytometer acquires three frames per second. Each of these frames contains the holograms of all the microscopic objects in a 24 µL sample volume directly above the image sensor at that time point during the flow. This full field-of-view raw frame undergoes several processing steps to localize and reconstruct the particles present in the liquid volume above the image sensor. First, the static objects (dust, etc.) in the field of view are digitally removed using a background subtraction algorithm and only the holograms of the objects located inside the continuously flowing water sample are preserved. This background subtraction algorithm calculates the background by calculating the average of the preceding 20 frames captured by the device. To eliminate static objects from a newly acquired frame, the background frame is first subtracted, and then the mean value of the background frame for each corresponding color channel is added back to preserve the image quality. The result is a full field-of-view image where only the diffraction patterns of the objects that move with the flow are present. The center coordinates of each object's hologram in the background subtracted frame is located by performing circular Hough transform on the image. For each of the located objects a 512×512-pixel region of interest, containing the hologram of the corresponding particle at the center, is cropped from the full field-of-view background subtracted image. Next, the red, green, and blue color holograms are extracted from these segmented Bayer patterned images, also removing the color crosstalk between the color channels. Since the color image sensor's Bayer pattern contains twice as many green pixels than red or blue pixels, the individual monochromatic holograms are resampled to generate a uniform regular sampling grid in each one of the color channels.

Here we did not interpolate in the green pixel grid, instead used a resampling process where the image gets rotated by 45° to maximize the quality of the reconstruction[26]. To obtain the three dimensional location of each one of the segmented objects, we automatically determine the reconstruction height by using the edge sparsity of the field's complex gradient in the blue color channel as an autofocusing metric[27]. The algorithm stores the three dimensional location of all the detected particles in previous frames, estimates their current location based on the flow profile inside the channel, and compares them to the newly detected particles' coordinates from the currently processed frame in order to avoid reconstructing or counting the same object twice during its flow through the device. At this point the list of newly detected objects in the frame and their corresponding three dimensional coordinates are determined. To create high resolution reconstructions of these detected objects of interest, the corresponding holograms are upsampled by a factor of four and then all the three color channels are propagated corresponding to their incidence angle and wavelength to the previously determined reconstruction distance using the angular spectrum method[13,28]. This process results in phase and intensity images at three different wavelengths, each with a size of 1024×1024 pixels for the detected objects in the flow, which are then processed by a *Giardia* classifier network (detailed next).

*Convolutional neural network for Giardia cyst classification*

The classification of the detected objects was performed using a deep neural network following the Densenet-121 architecture[24], with two output classes (*Giardia* and non-*Giardia*). Before passing through the network, the centers of the phase and intensity images are each cropped to 256×256 pixels. The three color channels of both the intensity and phase images are concatenated into a single matrix with six channels. Each of these matrices is then normalized by dividing them by the median value of each channel, as this process will remove slight intensity differences in between subsequent measurements. In order to train the network, a softmax cross-entropy based loss function (*L*) is used. To penalize false positives, the loss for the non-*Giardia* images is weighted twice as heavily, i.e.:

$$L = - y_{c=Giardia} \log(p_{c=Giardia}) - 2\, y_{c=non\,Giardia} \log(p_{c=non\,Giardia}) \quad (1)$$

where $y_c$ is a binary one-hot vector encoding the true label (i.e., it has a value of one if the object belongs to that class and is zero otherwise). The $p_c$ variables represent the probability of a classification decision, and are calculated using the softmax function, i.e.:

$$p_c = \frac{\exp(z_c)}{\sum_{k=1}^{2} \exp(z_{c=k})} \quad (2)$$

where $z_c$ is the value output by the neural network for each class, $c = Giardia$ or $c = nonGiardia$.

The network further reduces the number of false positives by adding a bias to the classification. In order for an object to be classified as a *Giardia* cyst, the output of the neural network for the *Giardia* class has to be higher than the output of the non-*Giardia* class by an empirically determined value of 2. Therefore, an object is only classified as a *Giardia* cyst when the following equation is satisfied:

$$z_{c=Giardia} > z_{c=non\,Giardia} + 2 \quad (3)$$

This network was trained using 243,774 images captured by our imaging flow cytometer, consisting of 131,215 *Giardia* images and 112,559 images coming from mixtures of other objects, classified as "non-*Giardia*". 80% of the images were used for training while the remaining 20% of these images were used for validation. The network training was implemented in Python version 3.6.2 using Pytorch version 1.2. The trained network was subsequently implemented in CUDA for real-time processing using a laptop computer (MSI GE75 with an Intel Core i7-8750H CPU, 16 GB of RAM and an Nvidia RTX 2080 GPU). The network was trained on a Windows computer using an Nvidia GTX 1080 Ti GPU.

*Graphical User Interface (GUI)*

The user operates the imaging flow cytometer through a custom developed GUI (see Figure 2). The user is able to set the liquid flow speed, the illumination pulse duration, the brightness of each LED, and the gain of the image sensor. All the image processing steps are done in real-time as the algorithm is running on the laptop's GPU. The GUI provides instantaneous visual feedback on the density of the objects present in the full field of view, and also shows, in real-time, reconstructed intensity images of the objects. The user can observe the size histogram of the detected particles and the number of *Giardia* cysts identified in the sample since the start of the measurement in real-time. The software is also capable of processing and analyzing a previously saved raw sensor image dataset.

*Water sample testing*

Flow cytometer enumerated and irradiated 1 mL *Giardia lamblia* cyst suspensions (Waterborne Inc.) were used for preparing the spiked water samples. The concentrations of these suspensions were 10, 50, 100, and 500 cysts in 1 mL of reagent grade water according to the manufacturer. The cysts suspensions were stored at a temperature of 6°C ± 2°C until the experiments were performed. Prior to the measurement, the cyst suspensions were brought to room temperature. The cyst suspensions were vortex mixed for 2 min and poured into a 50 mL disposable centrifugation tube (05-539-9LC, Fisher Scientific) containing 40 mL of sterile-filtered reagent grade water (FB12566506 or SE1M179M6, Fisher Scientific). To improve the recovery rate from the manufacturer's vial, 1 mL of sterile-filtered reagent grade water (23-249-581, Fisher Scientific) was added to the emptied 1mL tube which was vortexed mixed again for 15 seconds and then the contents were poured into the same 50 mL tube. This last step was repeated 8 times per sample to improve the recovery of cysts from the manufacturer's vial. Finally, the tube was filled up to 50 mL using filtered reagent grade water.

We used the same protocol to prepare *G. lamblia* spiked sterile-filtered seawater samples. Here, instead of sterile-filtered reagent grade water, we used filtered seawater (SEABH20L, Bigelow National Center for Marine Algae and Microbiota) and sterilized it using a disposable filtration unit.

*Preparation of 50 mL concentrated samples from 1 US gallon bottled water samples*

Each gallon of bottled water (Spring water, Ralphs) was on purpose contaminated with four vials (each 1mL) of *G. lamblia* cyst suspensions (Waterborne Inc.), two vials with a nominal cyst count of 500 and two others with a manufacturer provided cyst count of 100. A glass vacuum filtration unit (XX1019022, Fisher Scientific) was equipped with a filter membrane which had a pore size of 0.45 µm (1270061, Sterlitech

Corporation). The spiked 1-gallon sample was then poured into a glass vessel and filtered at a pressure of 50 kPa. The 1-gallon plastic bottle was rinsed twice using 250 mL of sterile-filtered reagent grade water and the rinse water was poured into the glass vessel for filtration. The filter membrane containing all the particles from the sample was removed from the filtration unit and placed into a sterile Petri-dish. ~15 mL of 0.05% [v/v] Tween®20 in sterile-filtered reagent grade water was added to the Petri-dish. Using a cell scraper (B002VBW7H0, Amazon Inc.) the particles captured on the membrane were released into the solution that was then collected in a 50 mL Falcon tube. This procedure was repeated 3 times per filter membrane to improve the particle recovery rate and to yield a final volume of ~50 mL concentrated water sample.

*Measurement workflow*

At the start of each measurement a new set of disposable components was installed into the flow cytometer to avoid cross-contamination between the samples. In case the effluent was used for subsequent fluorescence microscope based analysis (for comparison purposes), the output tube of the system was placed over a vacuum filtration unit containing a 0.45 µm pore size filter membrane (1270061, Sterlitech Corporation). Then, the flow cytometer was turned on and the controlling software was initialized. To reduce the amount of *Giardia* cyst adhesion within the inner walls of the disposable components (e.g., plastic tubing and sample channel) we have flown the washing buffer for 5 minutes at a flow rate of 100 mL/h. After surface activation, the inlet tube was placed from the washing buffer to the 50 mL tube containing the sample and the image acquisition and real-time analysis started using the GUI. The imaging flow cytometer automatically images and analyzes the 50 mL sample in 30 min using a laptop computer, counts the number of *Giardia* and non-*Giardia* particles and sorts the images of the objects identified as *Giardia* from the images of the rest of the detected micro-objects.

*Comparative fluorescence microscopy analysis of the effluent*

To perform comparative analysis for validation of our results, fluorescent labeling of the cysts captured on the filter membrane was performed using fluorescein labelled *Giardia* specific antibodies (A300FLR-20X, Waterborne Inc.). The stain was diluted to 1× according to the manufacturer's instructions. 200 µL of dye suspension was spread on the filter membrane containing the filtered effluent and was incubated at 37°C for at least 30 minutes for labeling the *G. lamblia* cysts. Manual counting of cysts captured on the filter membrane was performed using a FITC filter set and 20× to 40× objective lenses using a benchtop fluorescence microscope (Olympus BX51). The analysis of the membranes was completed within 24 hours after staining. Each membrane was manually scanned in a systematic snake scanning pattern to avoid missing any *Giardia* cysts. These results were then compared against label-free classification and quantification results obtained with our imaging flow cytometer.

**Conclusions**

We demonstrated a cost effective, field-portable and label-free, computational imaging flow cytometer that can perform in real-time automated detection and quantification of *Giardia lamblia* cysts in fresh and seawater samples. The system uses lensfree holographic microscopy with pulsed illumination to obtain in-focus phase and intensity images of the microscopic particles continuously flowing in a channel and subsequently classifies them as *Giardia* or non-*Giardia* using a trained convolutional neural network. The flow cytometer analyzes the *Giardia* content of liquid samples in real-time at a throughput of 100 mL/h, and was demonstrated to detect *Giardia* contamination in water samples containing < 10 cysts per 50 mL. This label-free cytometry technology can be useful to analyze and detect *Giardia* cysts in drinking water samples in resource limited settings and to periodically monitor the integrity of filters in water purification systems.

**Acknowledgements**

The authors acknowledge the US Army Research Office (ARO, W56HZV-16-C-0122).

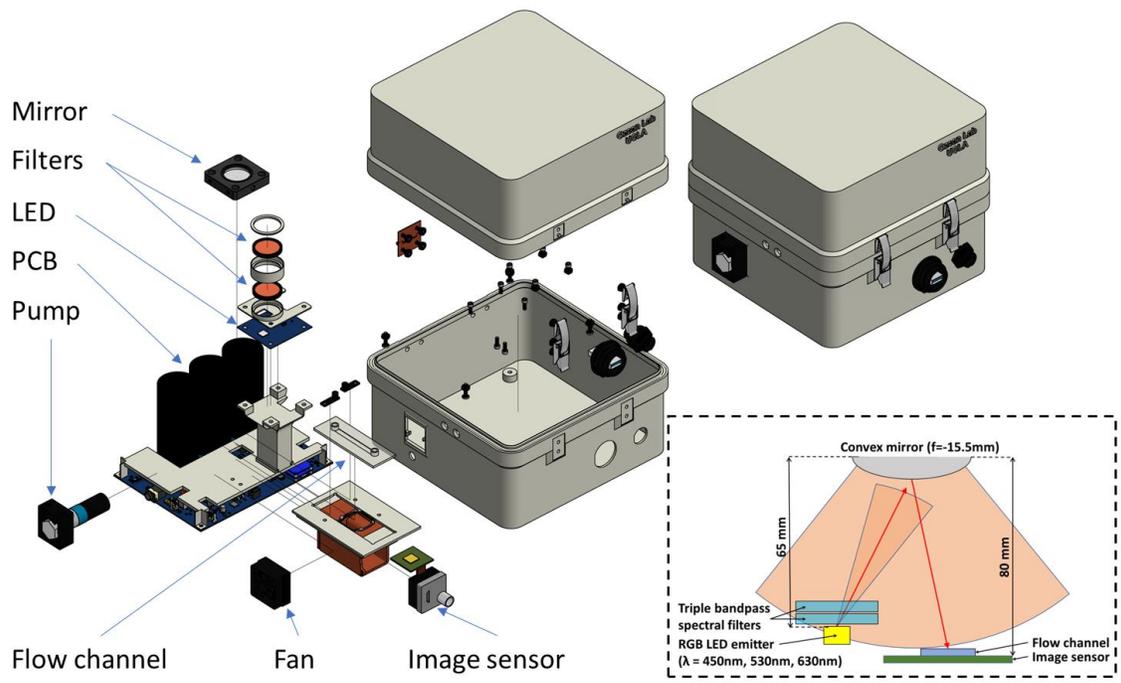
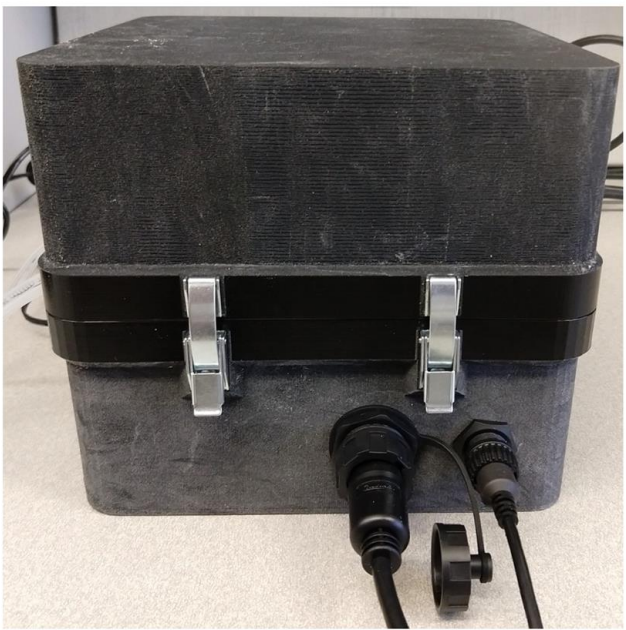
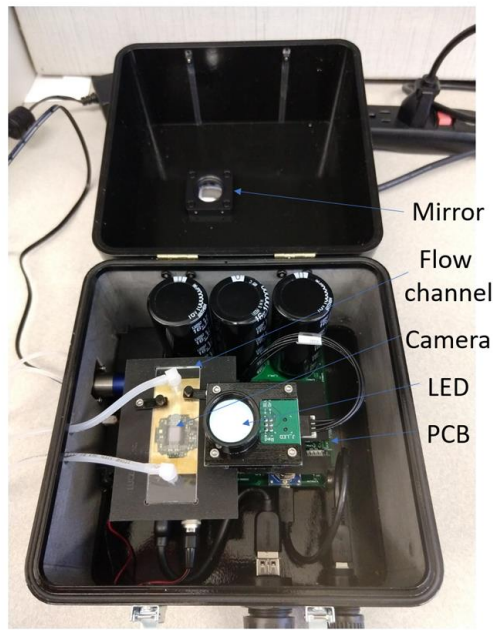

**Figure 1: Field-portable imaging flow cytometer for label-free detection and real-time quantification of *Giardia* cysts.** (Top) Schematic and exploded views of the holographic imaging flow cytometer. The inset shows a sketch of the optical system. The spatial and temporal coherence of the RGB LED are adjusted with bandpass filters and a convex mirror to provide pulsed illumination for color holographic imaging flow cytometry in a compact form factor. (Bottom) Photos showing the assembled device in closed and opened state.

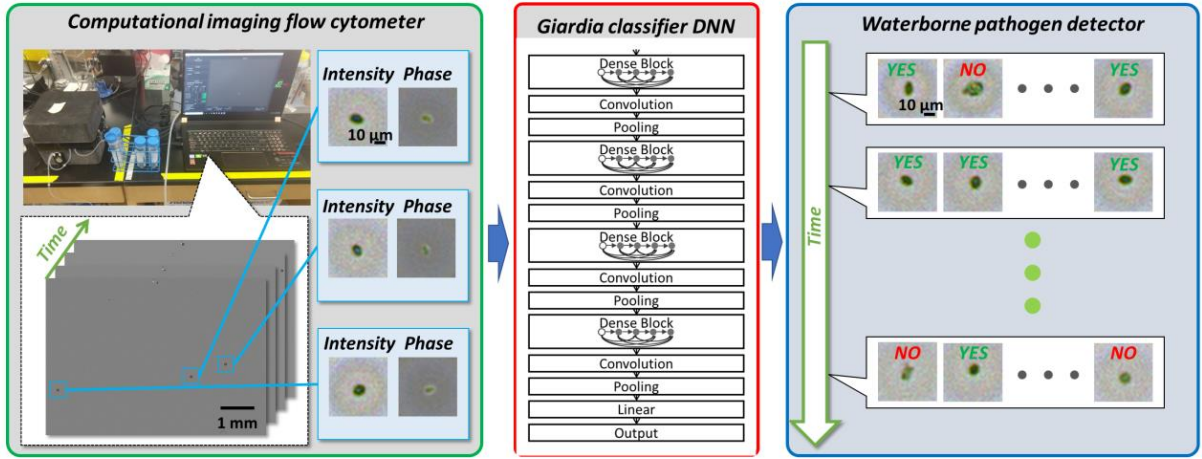

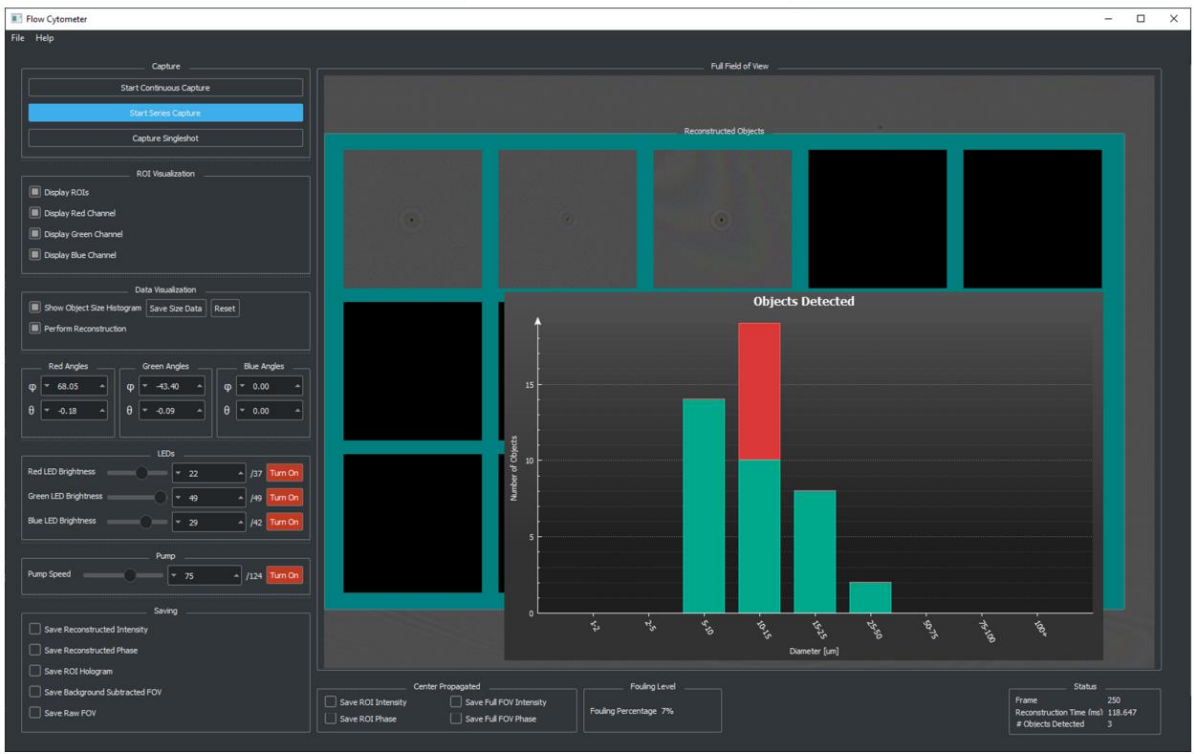

**Figure 2: Workflow of the field-portable imaging flow cytometer and a screenshot of the graphical user interface.** (Top) The flow cytometer screens the continuously flowing water sample for *Giardia* cysts at a flowrate of 100 mL/h by capturing lensfree holograms of microscopic objects inside the fluidic channel. Each frame has a sample field-of-view of ~30 mm$^2$ and corresponds to a liquid volume of 24 µL at a given time-stamp. The pulsed holograms of individual objects are segmented and reconstructed automatically. The measured phase and intensity information of all the objects are processed by a trained neural network to automatically identify *Giardia Lamblia* cysts in real-time using a laptop computer. During each measurement, the GUI (bottom image) displays a full field-of-view image, which allows the user to track the contents of the fluidic channel. The user can enable to view the reconstructed intensity images of the objects detected on the current frame. The user can also enable to view a size histogram of the objects

detected in the sample since the start of the measurement. On this histogram, the particles classified as *Giardia* cysts are shown in red color, while the non-*Giardia* particles are shown in green.

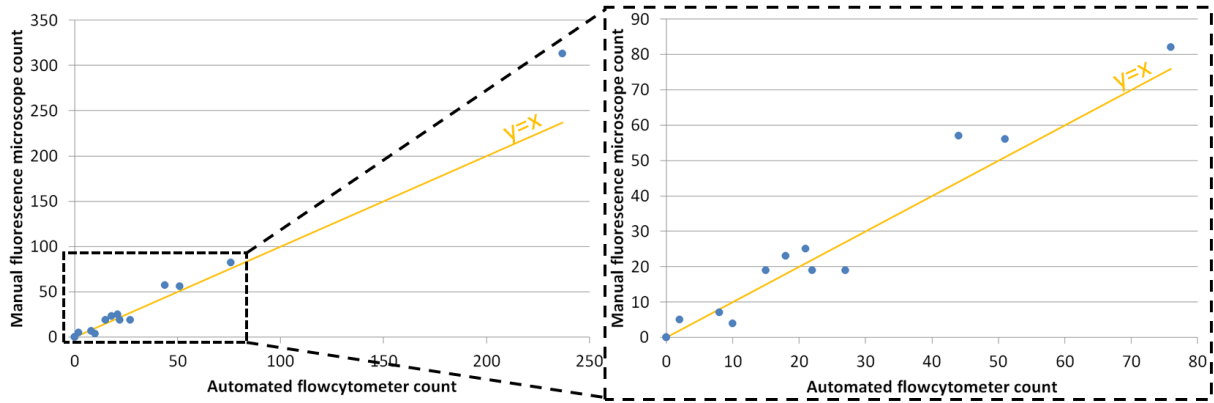

**Figure 3: Measurement results for 50 mL reagent water samples spiked with *Giardia lamblia* cysts.** For the samples containing over 20 cysts, the flow cytometer achieved a recovery rate of ~83% with a standard deviation of ~7%. For lower concentration samples containing less than 20 cysts the standard deviation of the measurements was ~4 cysts. The flow cytometer was able to correctly identify all 12 different water samples containing *Giardia lamblia* cysts as positive.

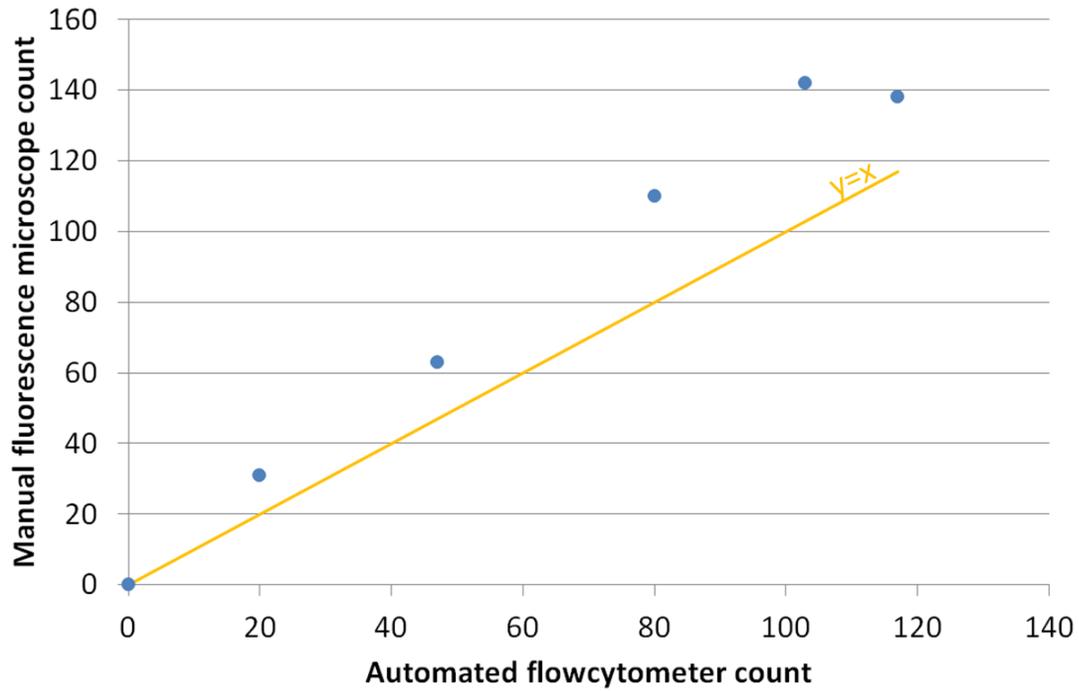

**Figure 4: Measurement results for 50 mL seawater samples spiked with *Giardia lamblia* cysts.** The flow cytometer achieved a recovery rate of ~74% with a standard deviation of ~6%. The flow cytometer was able to correctly identify all the 5 different seawater samples containing *Giardia lamblia* cysts as positive.

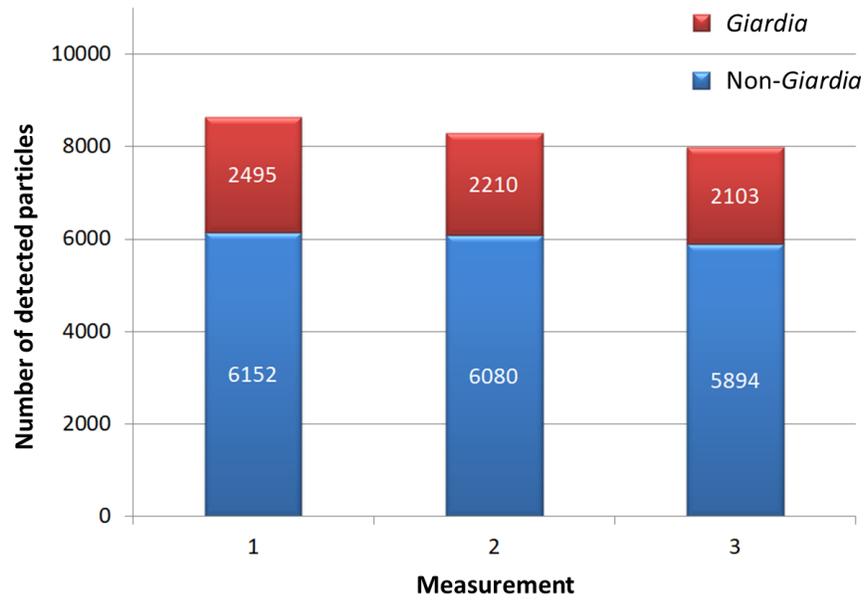

**Figure 5: Repeatability of the imaging flow cytometer.** A 50 mL sample was spiked with *Giardia lamblia* cysts and was measured three times in succession by collecting the effluent from the system and testing it again using the same device. The number of particles is slightly decreasing in between successive measurements due to the adhesion of the particles to the sample holder. This effect is more pronounced for the *Giardia* cysts, most likely due to their preferred adhesion to hydrophobic surfaces of our measurement system.

|  | *Giardia* cysts count ||  Number of Non-*Giardia* particles | Prediction |
|---|---|---|---|---|
|  | **Fluorescence microscope** | **Flow cytometer** |  |  |
| **Sample 1** | **126** | **97** | **23253** | **Positive** |
| **Sample 2** | **70** | **38** | **20902** | **Positive** |
| **Sample 3** | **139** | **31** | **25102** | **Positive** |
| **Control** | **0** | **0** | **14973** | **Negative** |

**Table 1: Experimental results on water samples (each 1 US gallon) spiked with *Giardia* cysts.** Each one of these 1 US gallon water samples was pre-filtered using a mechanical filter into a test volume of 50 mL before being analyzed by the portable imaging flow cytometer. All the positive samples as well as the control sample were correctly classified by the portable label-free cytometer.